\documentstyle[11pt,newpasp,twoside,epsf]{article}
\def\msun{M_{\odot}}
\def\gtorder{\mathrel{\raise.3ex\hbox{$>$}\mkern-14mu
             \lower0.6ex\hbox{$\sim$}}}
\def\ltorder{\mathrel{\raise.3ex\hbox{$<$}\mkern-14mu
             \lower0.6ex\hbox{$\sim$}}}

\def\edcomment#1{\iffalse\marginpar{\raggedright\sl#1\/}\else\relax\fi}
\reversemarginpar

\begin{document}

\title{Update on Pulsar B1620$-$26 in M4: Observations, Models, and Implications}

\author{Steinn Sigurdsson }
\affil{525 Davey Laboratory, Department of Astronomy \& Astrophysics,
    Pennsylvania State University, PA 16802}
\author{Stephen E. Thorsett}
\affil{Department of Astronomy \& Astrophysics, University of California, Santa Cruz, CA 95064}

\begin{abstract}
The combination of ever more precise radio timing data and serendipitous  {\it HST} 
data has confirmed that the outer companion to PSR B1620$-$26 is a planet.
Here we summarize the observational situation, including preliminary new timing solutions and the implications of the
measured system parameters. We detail the proposed formation scenarios, discussing the advantages and problems of each for explaining the origin of the triple, and we speculate on
some of the implications
for planet formation in the early universe. Future data on this system
will provide additional constraints on fundamental modes of planet formation.
We predict that many more exchanged planets will be discovered orbiting 
recycled pulsars in globular clusters as the sensitivity and duration of
radio timing increases. Strong observational tests of some of the alternative
formation models should be possible with additional data.
\end{abstract}


\section{Historical Background}
%{\bf 1. Background}\\

The long series of radio timing data of the triple pulsar PSR~B1620$-$26 (Lyne et al.\ 1988)
has now confirmed
that the second companion (Backer et al.\ 1993), is a $1-3 \, M_J$ substellar object---a planet---in a low eccentricity, wide
circumbinary orbit about the inner pulsar--white dwarf binary. The outer orbit
is significantly inclined to the inner orbital plane, and HST photometric data
confirms that the white dwarf is young, and is a proper motion member of the cluster
(Thorsett et al.\ 1999; Sigurdsson et al.\ 2003). 

The mass of the object is very well constrained. The timing data puts
a firm {\it lower} mass limit of a jupiter mass; as more data has come in,
the {\it upper bound} on the mass has steadily shrunk. When timing first hinted
at the presence of the third object in the system, the mass was only weakly 
constrained (c.f.\ Michel 1995), although various arguments suggested that a low
mass solution might be preferred (c.f.\ Thorsett et al.\ 1993, Sigurdsson 1993, Phinney 1993,
Rasio 1994, Joshi and Rasio 1997). 
By 1999 Thorsett et al., using measurements of the precession of the inner orbit produced by the tidal field of the outer object, had effectively excluded stellar mass companions,
and the main issue was whether the object was properly a brown dwarf, formed through
star formation processes, or whether it was a planet. When the assumption is made that the pulsar spin-period derivative is negligibly small (which is true if the magnetic field of the pulsar is a typical $3\times10^8$~G, Thorsett et al.\ 1993, 1999), then preliminary analysis of the most recent timing data finds a unique Keplerian solution, with $m_3\sin i\sim1.7M_J$, $e\sim0.13$, and orbital period $P_b\sim68$~yrs. (A detailed analysis, relaxing the assumption on the spin-period derivative and including orbital perturbation measurements to constrain $\sin i$ is in preparation.)

The serendipitous observations of the white dwarf companion in deep multi-epoch HST
imaging of M4 (Sigurdsson et al.\ 2003, Richer et al.\ 2002, 2003, Bassa et al.\ 2003) 
provide constraints on the
white dwarf mass and hence the inner binary inclination. These results, which confirm the prediction from the radio data, strongly favor a $1-3 \, M_J$ mass second companion.

We must ask ``is it really a planet?'' By definition, an object of this mass,
orbiting a star, would be a jovian planet if observed in the solar neighbourhood.
By inference, it is a gas giant. It seems very unlikely that a rocky or icy planet
of such mass could form in M4 in any circumstances.  A serious issue raised
by the existence of this object is whether it formed through the canonical
``bottom-up'' core accretion process, or whether it formed through
``top-down'' collapse of cold gas that had become secularly unstable
under its own gravity (c.f.\ Boss 2002, 1997).
The former process would imply the existence of lower-mass planets that had
failed to grow to the point where nebular gas accretes to the planet, and
hence we would predict the presence or rocky or icy terrestrial planets
in M4 specifically and in globular clusters more generally. In view of the
observed correlation between the incidence rate of observable ``hot'' and ``warm'' jovians
and host star metallicity in the solar neighbourhood (c.f.\ Fischer and Valenti 2003),
the possible prevalence of ``normal'' jovians around pop II stars would be of particular
interest, independent of the formation mechanism. Clearly the possibility of 
low mass planet formation around pop II stars has significant anthropic implications
and for considerations of the prospect for extraterrestrial life.
Conversely, if the planet formed through top down collapse, we can further ask whether it formed
in a cold protoplanetary disk, or as an independent quasi-spheroidal collapse---essentially as a low mass or failed brown dwarf. In principle, the different scenarios
are testable. The possibility that the difficult of growing a massive icy core
in low metallicity protoplanetary disks might be partially offset by the onset
of gas accretion at a lower core mass is particularly intriguing 
(c.f.\ Rice and Armitage 2003). 

In addition to the very precise radio timing measurements of the spin frequency
and its various derivatives and the observed time variation of some
of the orbital parameters, a number of other observables of the system
combine to provide key hints as to the formation and dynamical history of
the system. These almost completely constrain its possible past. The data allow
us to invert the dynamical history of the system with remarkable confidence
to recover the initial conditions with relatively little ambiguity:

\begin{itemize}
\item{} Almost as soon as the system was detected, the white dwarf orbital eccentricity
was noted to be anomalously large (McKenna and Lyne 1988). If the pulsar were spun-up
through conservative mass transfer from the progenitor of the white dwarf (c.f.\ Rappaport
et al.\ 1995, Phinney and Kulkarni 1994),
we would expect the white dwarf eccentricity to have been several orders
of magnitude smaller after the white dwarf detached. It is generally agreed
that the current eccentricity was dynamically induced after mass-transfer
was complete. The process by which this occurred has been somewhat contentious,
with a number of less than satisfactory processes proposed. 
The cause of the white dwarf orbital eccentricity is now almost certainly 
confirmed to be the Kozai mechanism (Ford et al.\ 2000), in which angular
momentum is exchanged from the inclination
%XXX 
of the planet's orbit to the eccentricity of the white dwarf's orbit.
The current orbits are known to be significantly non-coplanar (relative inclination
of about $40^{\circ}$), and the inclination must have been significantly larger still
before angular momentum exchange took place ($\sim 70-80^{\circ}$), for the Kozai
mechanism to have induced the observed white dwarf orbital eccentricity.
\item{} The inferred current and original high relative inclination between the
inner binary plane and the orbital plane of the planet is highly significant and
a major clue to the possible formation paths for this system.
\item{} The planet orbit is now thought to have relatively {\it low} orbital eccentricity.
This is a major constraint on formation scenarios, since most mechanisms that
lead to high orbital inclination naturally also lead to high orbital eccentricity.
Exchange processes that leave a planet in a stable circumbinary orbit tend
to naturally lead to moderate eccentricities and high inclination, since high
eccentricity post-exchange orbits are generally dynamically unstable,
and low eccentricity orbits are improbable due to the small 
available phase space at low eccentricity.
The current low planet orbital eccentricity may then be due to circularization of the initially
moderately eccentric planet orbit during adiabatic mass loss of the white dwarf
progenitor envelope, during which the system mass went from an initial $\sim 2.3 \msun$
to the current $\sim 1.8 \msun$. The planet's semi-major axis increased in
proportion at that time.  This sequence requires the planet be in place before
the white dwarf progenitor evolved off the main-sequence. 

%XXXX
To a good approximation, during the RGB evolution of the white dwarf progenitor, we can
treat the planet as a test particle in a Keplerian orbit about a point mass potential
with a central mass equal to the total mass of the neutron star and its companion.
The loss of the envelope mass is slow, and the planet orbital evolution is adiabatic; its
energy changes in response to the loss of mass from the center, but the other integrals of motion
are invariant. Eccentricity in general is an orbital parameter, not an invariant,
however, for a Keplerian orbit, $1 - e^{2}_{p} \propto  E_p J^2_p$; the eccentricity
is a function of the integrals only, and is invariant to isotropic slow mass loss.
The current eccentricity, $e_{fin} = 0.2 \pm 0.1$, from current fits to the timing data, is somewhat
lower than expected from exchange models (which predict $e \sim 0.3-0.7$).
It is possible this system had an unusually low initial post-exchange eccentricity, but
this then raises the same fine tuning problems present in other scenarios.

Alternatively we can postulate a post-exchange planet eccentricity of about $0.5$ and ask
whether circularization could have taken place without substantial decrease in inclination.
Given a current semi-major axis of $\sim 20 AU$, the post-exchange, pre-RGB phase, semi-major axis
must have been $\sim 16 AU$, with a corresponding periastron of $\sim 8 AU$, assuming a median post-exchange
eccentricity of $0.5$.
As the white dwarf progenitor evolved, its orbit initially circularized, and then expanded
as conservative mass transfer took place, until it reached its current orbit; the time scale
for such a mass transfer phase and spin-up is $\gtorder 10^8$ years.
With a periastron of $\sim 8 AU$, compared with a white dwarf progenitor final orbit of $1 AU$,
we find that substantial tidal circularization of the planets orbit could have occurred at late stages of 
the mass transfer phase, yet the time scales are long enough that complete circularization is precluded.
Following Verbunt and Phinney (1995), we estimate $\ln \Delta e \sim -1$, or $\delta e \sim 0.3$, 
consistent with an initial post-exchange orbital eccentricity of $\sim 0.3-0.5$, in the range predicted,
and consistent with the currently observed eccentricity.  Inclination changes during this phase should have been
negligible; unless there was significant planar outflow during mass transfer, in which case coupling
of the planet to the excretion ring during plane crossings could in principle be a concern.
The system here of course is a triple, not a point mass secondary interacting with an extended massive central star,
but the additional lever arm of the giant rotating about the inner system center-of-mass
will in general somewhat enhance the tidal torque, making this scenarios more plausible.
The inferred eccentricity of the planet's orbit, before the evolution of the pulsar companion
to white dwarf, is then about $0.5$, exactly in the range predicted for an exchange scenario;
and is not sensitive to the exact value of the current eccentricity. Allowing the neutron star
to accrete $\sim 0.1 \msun$ during mass transfer,
with a correspondingly larger final total central mass, does not significantly
change the predicted initial eccentricity; if there was slow isotropic mass loss, some additional
circularization due to tidal interaction with the wind is conceivable. Mass accretion by the planet
is negligible for all scenarios; mass loss due to LMXB ablation of the planet is also negligble
for plausible LMXB phase luminosities.

If the planet eccentricity was initially low, negligible tidal circularisation took place,
if the planet eccentricity was initially substantially higher, then there was opportunity for
significant tidal circularisation during the mass transfer phase; if the mass transfer phase was extended.
Since this is a long period, low mass binary pulsar, we expect sub-Eddington mass transfer with
mass transfer time scales of $O(10^8)$ yrs or longer, depending on the core mass of the main sequence
progenitor post-exchange and the eccentricity of the main sequence progenitor orbit before circularization
of the inner orbit and onset of mass transfer (cf Burderi et al 1996, Webbink et al 1983).

%XXXXX

\item{} The HST data reveals the white dwarf is young. The main sequence
progenitor was presumably as old as the cluster, 12.7 Gyrs, but the white dwarf
emerged from the mass transfer phase just under 0.5 Gyrs ago. This is consistent
with a single exchange, where the white dwarf progenitor was acquired by the neutron
star some 1--2~Gyr ago, and the recoil induced by the super-elastic exchange
ejected the system to the outer parts of the cluster, where stellar densities
are low and interaction time scales are long. Therefore the system has probably
been mostly dynamically isolated since the exchange.
\item{} The projected pulsar position is somewhat outside cluster core. The true 
position of the pulsar is of course probably somewhat further from the core than
we see in projection. Since the system is significantly more massive than the main sequence turnoff,
or indeed any likely other stellar component in M4, it tends to sink to the core
through dynamical friction. The current characteristic time-scale for the orbit of the
triple to sink back into the cluster core through dynamical friction is about
a Gyr, and its current position is consistent with an initial orbit with an apocenter
beyond the half mass radius and total dynamical lifetime of 2-3 Gyrs, which is consistent
with a single exchange origin, combined with ejection from the core.
\end{itemize}
Given these data, and the current pulsar kinematics, we can construct a canonical
formation scenario for the system.

%{\bf 2. Canonical Formation Scenario}\\
\section{The Canonical Formation Scenario}

The pulsar is very unlikely to have formed as a member of the current binary.
A small fraction of neutron stars appear to have sub-solar mass companions.
For the white dwarf to be at the current 1 AU orbit, its progenitor must
have been in a tighter orbit initially, and for such a system to have
survived the supernova that formed the neutron star, there must have
been a modest natal kick on the system.
However, given a kick that would leave a tight low mass binary, the system 
most likely would have been ejected from the cluster by the net kick on the center
of mass of the system (c.f.\ Phinney and Kulkarni 1994).

More likely the pulsar was originally a member of a pulsar-heavy white dwarf system, 
descended from intermediate mass binaries (Davies and Hansen 1998),
which is consistent with retention of the system in the globular cluster at formation.
The pulsar would then have started as a normal slow pulsar, and been recycled
to millisecond periods by its original companion (most likely $\sim 0.7 \msun$ white dwarf
in an orbit with semi-major axis of order $0.3 AU$). 
The system formed with the cluster, and spent most of the subsequent time
in the cluster core. M4 presumably has been slowly evolving in density, with
the core density increasing somewhat in the last few Gyrs as the cluster
evolves towards core-collapse (c.f.\ Meylan and Heggie 1997). 
Then, some 1-2 Gyrs ago, the neutron star binary
encountered a main-sequence star near the turnoff mass. Given the structure of
the cluster, the most likely star to undergo an exchange with a neutron star
binary is a turnoff mass star (Sigurdsson 1993b, Sigurdsson and Phinney 1995).
That this happened relatively recently in the cluster history is not coincidental, the pulsar is 
bright and therefore likely relatively recently recycled, and given cluster
dynamical evolution, the probability of an exchange taking place has increased
with time over the last few Gyrs. {\it The original white dwarf member of the binary
is ejected from the system. The main sequence star, with a mass somewhat higher than
the original white dwarf, becomes a member of the binary. The current white dwarf
member of the binary is the descendant of this main sequence star}. 

Such an exchange is super-elastic, and the system recoils from the core
of the cluster with a substantial velocity, traveling far from the core
(Sigurdsson 1993a, 1995). Mass-transfer ensues when the new member of the system
evolves off the main sequence, the orbit circularizes and expands, and
the system is x-ray luminous. The current pulsar, recycled for the second time,
emerges when the white dwarf forms and the system detaches. The white dwarf
is under-massive for its progenitor mass, as observed, since the RGB phase 
was terminated prematurely, and
no helium core burning took place (Sigurdsson et al. 2003, Rappaport et al. 1995).

In the particular scenario proposed by Sigurdsson (1993a, 1995) the main sequence
star was a primordial cluster star, formed in the outer part of the cluster, and it spent
most of its life in the outskirts of M4,  entering the core relatively recently. 
This is not improbable. Half the stellar mass of the cluster is outside the half-mass
radius (which does not vary much during the dynamical evolution of the cluster),
and at late times, as the turnoff mass of the cluster approached that of this particular star, 
the white dwarf progenitor would have sunk to the core through dynamical friction, 
because the turnoff mass
is higher than the mean stellar mass.
The planet was formed around the star at the time the cluster formed; it was
possibly one of many planets in the system (most likely the most massive planet),
and it exchanged into the current configuration during the same encounter that
the white dwarf progenitor, its parent star, was exchanged into the system 
(Sigurdsson 1992, 1993a,  Ford et al.\ 2000). The probability of a planet
exchanging to be bound to the neutron star at the same time as the main sequence
star is about 10-20\% ({\it ibid}). If there was more than one planet in the system, the
exchange probability is increased in proportion, to first order; 
if more the one planet is exchanged, the final state
is generally unstable and there are subsequent ejections or collisions
until only one planet remains. Although there is ``room'' in orbital space
for multiple planets, the probability of more than one remaining is low
for any given system.
As note above, this scenario predicts high inclination, and a modest initial 
eccentricity, $e_{in} \sim 0.3-0.7$ for the planet's orbit, somewhat higher than currently observed
(Sigurdsson 1993a). 

The system is unstable to subsequent encounters, but while it is outside
the cluster core, the probability of another encounter is small, and the
expected lifetime of the system is many Gyrs. Once the system returns to the core,
the lifetime of the planet to disruption by another encounter is $O(10^8)$ yrs.

If correct, the proposed scenario has significant implications. 
{\it A priori}, observing an exchanged planet around one of the first discovered
and longest observed globular cluster pulsars is likely only if planets are prevalent
around globular cluster main sequence stars.
Planet formation in dense environments is thought to be inhibited both dynamically and
because of the ionizing radiation from hot, young cluster member stars (c.f.\ Bally 2003),
although recent work shows the inner protoplanetary disk may be robust and generally survive
long enough for planet formation in most of the cluster volume (Bally NAI AbSciCon 2004 meeting),
in which case most stars formed in globular clusters may have had the opportunity for planet
formation to proceed in any protoplanetary circumstellar disks.

\subsection{Alternative Scenarios}

\begin{itemize}

\item{}The most straightforward alternative is a double exchange scenario (c.f.\ Joshi and Rasio 1997).
In this scenario, the pulsar-white dwarf system forms normally, as above, but the planet
is not initially present. Rather the planet is acquired through a second independent
exchange encounter outside the cluster core.
This scenario is attractive in that it removes constraints on the sequence of timing
of events, the pulsar recycling may have occurred at any point,  and the tight
restrictions on time scales for future disruption are removed since the planet
may have been acquired recently. In objection to this scenario, there is the observed
fact that the white dwarf formed recently, and precisely on the time scale required
to match the single exchange scenario; and, the low orbital eccentricity of the planet
suggests strongly that the orbit circularized after exchange, 
which requires the
planet to have been acquired before the white dwarf progenitor evolved off the 
main sequence (c.f.\ Bailyn et al.\ 1994).
In either case, the double exchange scenario also implies the presence of jovian planets
in wide orbits about cluster main sequence stars, and that these planets are not rare. 
Thus the implications for pop II planet formation hold.
\item{}An alternative scenario is that the exchange process proceeded as above,
but that the parent star of the planet was a galactic field interloper, a disk star
either passing through the cluster, or, more likely, capture by the cluster during disk passage
(c.f.\ Bica et al.\ 1997).  Exchange with a transient disk star is unlikely, because of the
high relative velocity and short time scale for passage on an unbound disk star.
A disk star captured by the globular during disk passage is more likely to have had an
opportunity for exchange.
In this case the planet formed around a metal rich pop I star
and implications for planet formation are less interesting.
Arguing against this scenario, is that although some models suggest some globulars
maybe significantly contaminated by disk interlopers, the M4 stellar population
is well studied, known to be homogenous and coeval, and proper motion membership
of stars has been established (c.f.\ Richer et al.\ 2002, Ivans et al.\ 2003).
Thus interlopers can not be common in M4, and the {\it a priori} probability
of a rare interloper undergoing an exchange encounter with the one pulsar is very low.
\item{}A commonly raised concern is whether the planet in this system could have formed
in situ after the supernova that presumably lead to the neutron star formation,
by analogy with the well known PSR~B1257+12 planet system (Wolszczan and Frail 1992).
This is not possible. The planet is in a high angular momentum orbit, to transport
it there from a close low angular momentum orbit without disrupting the system
is very unlikely; to transport it to a high inclination, low eccentricity orbit
is a negligible probability process.
\item{}Livio et al.\ (1992) proposed that planet formation might be efficient in the
post-merger debris of a WD-WD merger. One might speculate that the PSR~1620-26 system
is a descendant of such a system. The primary objection here is the same as
for post-supernova formation; such a process would be expected to lead to compact low
angular momentum systems. 
\item{}A somewhat intriguing possibility is that the pulsar formed from accretion induced collapse
(AIC; Bailyn and Grindlay 1990).
In this scenario, the system formed from a high mass white dwarf which presumably collapsed
due to mass transfer from the progenitor of the young white dwarf (although one can
imagine combining an AIC scenario with an exchange process). Then the planet would either
be exchanged from another cluster star, as before; or formed in situ. If it formed
at the same time as the stars, then its survival in a circumbinary orbit is remarkable
and the system must have formed in the far outskirts of the cluster and not yet reached the
cluster core. An attractive aspect of the scenario is that the planet would have formed in
a circular orbit much closer to the binary, with the current orbit coming from adiabatic 
expansion during mass loss, and the planet eccentricity induced by the sudden loss in rest
mass due to neutrino flux at AIC. A strong counterargument to this scenario is that the planet
is in a high inclination orbit, and the white dwarf orbital eccentricity is too low for an AIC scenario:
it ought to be comparable to the planet orbit eccentricity. The eccentricity discrepancy can be
overcome if there was mass transfer after AIC, which terminated before re-circularization was complete,
which may require some fine tuning of time scales. 
\item{}Another alternative is formation during the post-main sequence evolution of the
white dwarf progenitor (c.f.\ Livio and Pringle 2003). In this scenario, metal rich outflow
through the outer Lagrange point forms an excretion disk around the neutron star-giant system,
where the material can cool and planet formation may proceed, in principle.
We note that this particular system never underwent an AGB phase, since core evolution was terminated
during RGB ascent. Outflows also occur during the early stages of RGB evolution,
but the material will not be significantly enriched in metals, since no helium core
burning has taken place.  Since one motivation for alternative planet formation 
scenarios is to avoid the difficulty of planet formation in cool, metal poor, protoplanetary disks,
it is presumably harder to form the same planet in a hot, metal poor, thick excretion disk. 
Note also that the planet then has to form around a low mass x-ray binary with strong
ambient ionizing radiation, and the formation must take place in the outer disk, unlike
the case of PSR~1257+12, where planet formation took place in the optically thick inner
disk (Miller and Hamilton 2001, Hansen 2002).

{\it Independent of theoretical arguments, the alternative scenario where the planet
formed in a hot RGB envelope post-main sequence is testable. Exchanged planets will only
be found around pulsars in dense, low dispersion environments, such as globular clusters.
Planets formed in RGB envelopes must also presumably form around metal rich RGB companions
of disk pulsars which are members of low mass binaries. Therefore this alternative scenario
predicts that low mass binary pulsars in the Galactic disk should have circumbinary jovian planets,
a possibility which is not allowed by the exchange scenario.} 
\end{itemize}

As always when constructing theoretical scenarios for unique observed astronomical systems,
one can imagine unique formation paths which can not be excluded by the present data.
Judicious, but not over-exuberant, application of Ockham's Razor is then probably the
best way of choosing favored scenarios.
We note that some of the alternative formation scenarios involve the essential
point of dynamical exchange of a pre-formed planet which was a member of a main 
sequence star system with
a primary star which was solar like but from the metal poor population II.
%\begin{figure}
%\plottwo{fig1.ps}{fig1.ps}
%%\centerline{\psfig{figure=fig1.ps,width=5.25in}}
%\caption{
%}
%\end{figure}
%\subsection{BH binaries in GCs}
%%{\bf 2.2 BH binaries in GCs}\\
%
%\noindent$\bullet$  

\section{Future Implications}

In the immediate future, further observations of the system will take place:
spectroscopy of the white dwarf, using Gemini, to confirm the photometric estimates of
mass, composition and cooling age.
If additional {\it HST} data are taken, then the 
proper motion of system will be measured to a high precision compared to the
internal cluster dispersion. There is little prospect of system radial velocity
measurements with current generation of instruments and telescopes, and obtaining the
true radial position of the pulsar in the cluster is not currently possible  (although measurements of the apparent orbital period derivative may eventually set useful constraints on the acceleration of the system in the mean cluster potential, which when coupled with a cluster model could yield the 3d pulsar position relative to the center).
However, the orbit of the system in the cluster potential will be significantly
constrained by proper motion studies and will test the formation scenario.

More importantly, PSR~B1620$-$26 is proof of concept that precision timing of
pulsars can find diverse planetary companions, and complements the breakthrough
discovery of the first planetary system around PSR~B1257+12 
(Blandford et al.\ 1987, Wolszczan and 
Frail 1992).
We need to find more pulsar planetary systems, and we should expect to do so as
more data becomes available. This pulsar is bright, nearby and has been
monitored for a long time, which maximizes the prospect for detecting
a planet in the system. The planet is also quite massive, which makes it
easier to detect.
This technique could find terrestrial planets if they exist, and as discussed 
in Sigurdsson (1992) it should be possible to distinguish natively formed planets
like those in PSR~1257+12 from planets formed around main sequence stars and 
exchanged into a pulsar orbit by the kinematics.
Any further detection would place important constraints
on planet formation. Detection of low mass planets in pop II systems
would show core accretion formation is efficient even at low metallicities.
Absence of low mass planets and a preponderance of super-jovians (modulo 
the selection bias towards detection of high mass planets) would provide
strong support for formation through disk instabilities.
It is of course also possible that both mechanisms operate
(c.f.\ Haghighipour and Boss 2003, Durisen et al.\ 2004).

If planet formation is prevalent around metal poor population II
stars, which is implied but not absolutely proven by the observation
of a single example, then there are significant anthropic implications,
for the prospects for origin and evolution of extraterrestrial life and SETI,
and for strategies for future planet searches. In particular, pop II planets
would be a strong test of both the ``Rare Earth'' hypothesis (Ward and Brownlee 2000),
and the conjecture of the ``Galactic Habitable Zone'' (Gonzalez et al.\ 2001). 

A further test of planet formation theory would come through testing of
planet migration scenarios and whether migration is a dominant process in
formation of planetary systems, or a fringe process made prominent by the
selection biases of early phases of planetary observational searches.
In contrast to the positive detection of a planet in M4, is the absence
of detection of short orbital period jovians in a transit search of 47 Tuc 
(Gilliland et al.\ 2000). 
A number of explanations of the discrepancy are possible: it could be random chance,
either the detection in M4, or the non-detections in 47 Tuc could be small number
statistics and moderate luck. Alternatively, globular clusters could be heterogeneous
with planets forming in some clusters, but not others (c.f.\ Soker 2003), which is
a testable hypothesis if many systems are found.
Another intriguing alternative is that migration is metallicity dependent and
the prevalence of observed planets around metal rich stars in the solar neighbourhood
reflect this. Livio and Pringle (2003) find a relatively weak theoretical dependence
of the migration process on the disk metallicity, but if migration is rare and not
dominant (i.e., if most nearby stars have long period jovians), then maybe we can
infer that migration is a ``touch'n'go'' process, and a small difference in the
physics results in a large difference in the incidence. Alternatively, we can conjecture
that migration is sensitive to external perturbations, dynamical or radiative, and 
therefore inhibited to some extent in dense star forming regions. 

In conclusion, the confirmation of the second companion of PSR~B1620$-$26 as a planet
is a potential key step in our understanding of planet formation processes and
a way-guide for planet detection strategies. As a unique example, our inferences
are necessarily weak and conditional, but if further analogous examples are found
we can make strong tests of theories of planet formation and scenarios for
the origin and evolution of life in the universe.

\acknowledgments
We thank Tom Abel, Brad Hansen and Harvey Richer for helpful discussions.
SS's research was supported in part by NSF grant PHY-0203046
and the Center for Gravitational Wave Physics, an NSF funded Physics Frontier Center
at Penn State, and by the Penn State Astrobiology Research Center, supported
by the NASA Astrobiology Institute.
SET thanks his many collaborators who share credit for the observations of this
system, particularly Zaven Arzoumanian, Ingrid Stairs, and Joe Taylor. His work is supported by NSF grant AST-0098343. We also thank the Aspen Center for Physics for its hospitality.

\end{document}